\documentclass[twocolumn,showpacs,showkeys,preprintnumbers,
               amsmath,amssymb,floatfix,10pt]{revtex4}
\usepackage{graphicx}
\usepackage{dcolumn}
\usepackage{bm}
\usepackage{graphicx}
\usepackage{subfigure}
\usepackage{amsmath}
\usepackage{amssymb}
\usepackage{wrapfig}
\usepackage{color}
\usepackage[english]{babel}
\usepackage[warning,math]{easyeqn}
\usepackage[thinlines]{easybmat}

\newcommand{\scs}{\scriptscriptstyle}

\def\g{\gamma}

\def\a{\alpha}

\def\<{\langle}
\def\>{\rangle}
\begin{document}
%
\preprint{}
%
\title{Discrete breathers in nonlinear
network models of proteins}
\author{B. Juanico}
\author{Y.--H. Sanejouand}
\affiliation{
Ecole Normale Sup\'erieure,
Laboratoire Joliot-Curie, USR 3010 du CNRS,
46, all\'ee d'Italie,   
69364 Lyon Cedex 07, France\email{Yves-Henri.Sanejouand@ens-lyon.fr}}
\author{F. Piazza} 
\author{P. De Los Rios}
\affiliation{Ecole Polytechnique F\'ed\'erale de Lausanne,
Laboratoire de Biophysique Statistique, ITP--SB,  
BSP-720, CH-1015 Lausanne, Switzerland}  
%
\begin{abstract} 
We introduce a topology-based nonlinear network model of protein dynamics with the aim 
of investigating the interplay of spatial disorder and  nonlinearity. We show that 
spontaneous localization of energy occurs generically and 
is a site-dependent process. Localized modes of nonlinear origin form spontaneously 
in the stiffest parts of the structure and display site-dependent activation energies. 
Our results provide a straightforward way for understanding 
the recently discovered link between protein local stiffness and enzymatic activity. 
They strongly suggest that nonlinear phenomena may play an important role in enzyme
function, allowing for energy storage during the catalytic process.
\end{abstract} 
%
%
\pacs{63.20.Pw; 87.15.-v; 46.40.-f}
%
%
%
%
\keywords{Discrete Breathers, Elastic Network Models, Normal Mode Analysis, Nonlinearity}
\maketitle
%
The predictions of elastic network models (ENMs) of
proteins~\cite{Tirion:96,Bahar:97,Hinsen:98,NMA} 
have proven useful in quantitatively describing amino-acid fluctuations 
at room temperature~\cite{Tirion:96},  
often in good agreement  with 
isotropic~\cite{Bahar:97}, as well as anisotropic measurements~\cite{Phillips:07,Maritan:02}.
Moreover, it has been shown
that a few low-frequency normal modes can provide fair insight on
the large amplitude motions of proteins 
upon ligand binding~\cite{Tama:01,Delarue:02,Gerstein:02}, as previously noticed when 
more detailed models were considered~\cite{Brooks:85,Marques:95,Perahia:95},
also by virtue of the robust character of the collective functional motions~\cite{Nicolay:06}.
 
However, low-frequency modes of proteins are known to be highly 
anharmonic~\cite{Levy:82,Go:95}, a property which has to be taken into account
in order to understand energy storage and transfer 
within their structure as a consequence of ligand binding, chemical reaction, 
{\it etc}~\cite{Straub:00,Leitner:01}.
Indeed, there is growing experimental evidence that  long-lived modes 
of nonlinear origin may exist in proteins~\cite{Edler:2004uq,Xie:2000fk}.
Likewise, many theoretical studies have appeared suggesting 
that localized vibrations  may play an active role in, e.g., 
enzyme catalysis~\cite{Sitnitsky:06}. 
These include topological excitations such as solitons~\cite{dOvidio:2005qy} as well
as discrete breathers (DBs)~\cite{Archilla:2002lr,Aubry:01}.

\begin{figure}[t!]
\vskip -1.8 cm
\includegraphics[width= 11.0 truecm,clip]{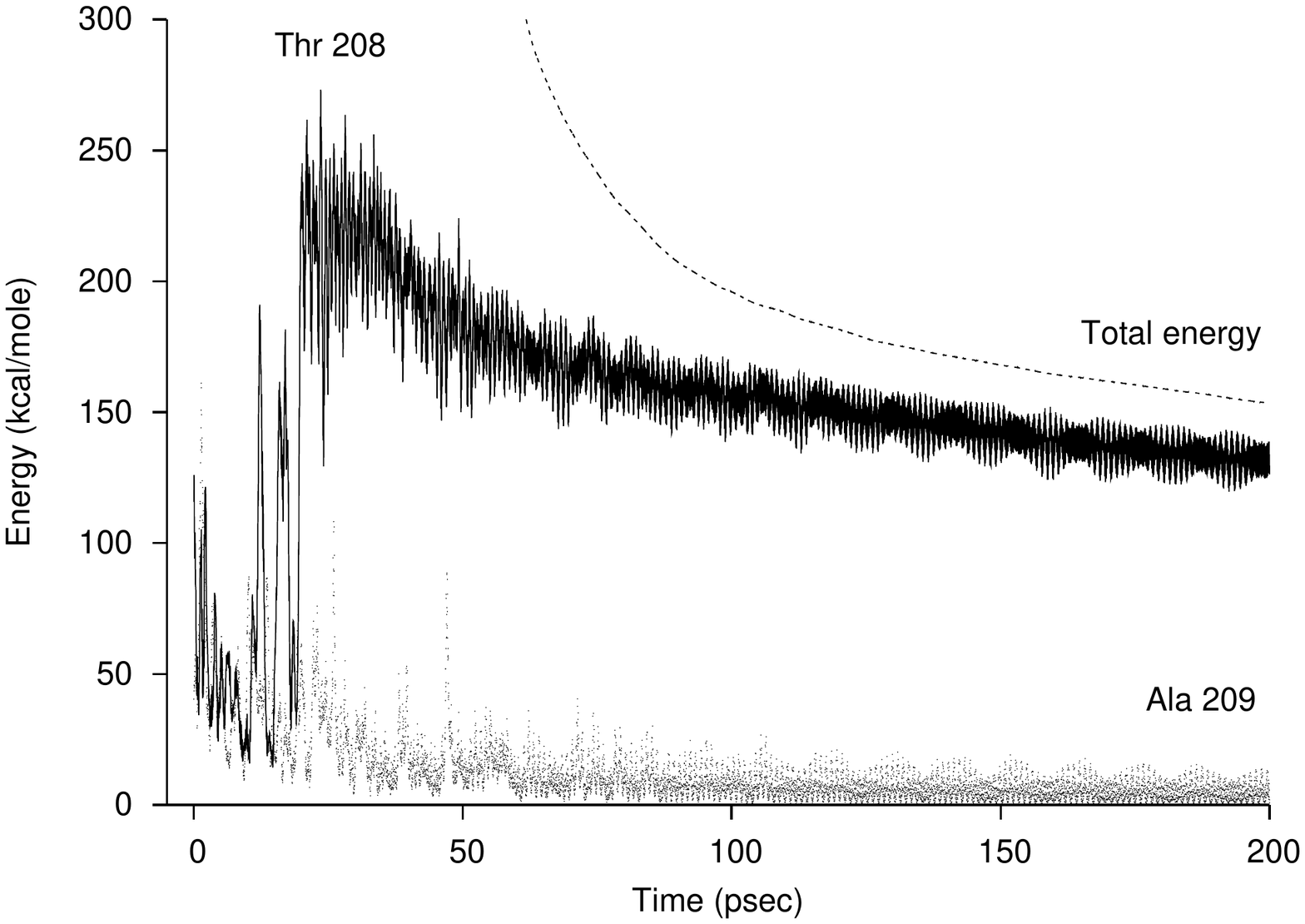}
\vskip -4.0 mm
\caption{\small \label{etot} 
Energy as a function of time, when citrate synthase is cooled down
as a consequence of surface friction.  Dashed line: total energy.
Solid line: energy of Threonine 208,
the amino-acid the most involved in the DB. 
Dotted line: energy of Alanine 209, also involved in the DB. 
$k_{\scs B} T_{eq}=20$ kcal/mol.
}
\vskip -5.0 mm
\end{figure}

The latter are nonlinear modes that 
emerge in many contexts as a result of both nonlinearity and discreteness~\cite{Flach:1998fj}.
Although their existence and stability properties are well understood in systems with 
translational invariance, much less is known of the subtle effects arising
from the interplay of spatial disorder and anharmonicity~\cite{DB:04,nonlin-disorder:01,Rasmussen:1999vn}.
For this purpose, in the present work we introduce the nonlinear network model (NNM).
Our aim is to extend the simple scheme of ENMs,
known to capture the topology-based features of protein dynamics~\cite{Tirion:96,Bahar:97,Hinsen:98}, 
by adding anharmonic terms.
Within the NNM framework, we show that spontaneous localization
of energy can occur in protein-like systems and that its properties may be
intuitively rationalized in the context of specific biological functions. 
In our model, the potential energy of a protein, $E_p$, has the following form:
\begin{equation}
\label{FPU}
  E_p=\sum_{d_{ij}^0 < R_c} \left[ 
                             \frac{k_{2}}{2} (d_{ij}-d_{ij}^0)^2 +
                             \frac{k_{4}}{4} (d_{ij}-d_{ij}^0)^4 
                            \right]
\end{equation}
where $d_{ij}$ is the distance between atoms $i$ and $j$, 
$d_{ij}^0$ their distance in the 
structure under examination (as e.g. solved through X-ray crystallography) 
and $R_c$ is a cutoff that specifies the interacting pairs.
As done in numerous studies, only C$_\a$ atoms are taken into account~\cite{NMA}
and $k_{2}$ is  assumed to be the same for all interacting atom pairs~\cite{Tirion:96}. 
As in previous ENM studies~\cite{Delarue:02,Helene:03}, we take $R_c=$10~\AA, and  
fix $k_{2}$ so that the low-frequency part of the linear spectrum match 
actual protein frequencies,
as calculated through realistic force fields~\cite{Brooks:85,Marques:95,Perahia:95}. 
This gives $k_{2}= 5$ kcal/mol/\AA$^2$, with the mass of each C$_\a$ fixed to 110 a.m.u., 
that is, the average mass of amino-acid residues.
Note that standard ENM corresponds to $k_{4} = 0$, 
while in the present work $k_{4} = 5$ kcal/mol/\AA$^4$.

Proteins live and perform their functions immersed in water and exchange energy 
with the solvent through their sizable surface portion. In a previous paper
we showed that complex energy relaxation patterns are  observed 
as a result of  the inhomogeneity of the coupling 
to the solvent of bulk and surface atoms~\cite{Piazza:05}. 
In the presence of nonlinearity, boundary relaxation is known to drive 
a wide array of systems towards regions of phase space corresponding
to localized modes that emerge spontaneously~\cite{Aubry:96,Piazza:03,Reigada:2003,Livi:2006vn}.
Thus, in order to study {\em typical} excitations of nonlinear origin in protein structures,
it appears natural to perform a boundary cooling experiment. 
Our protocol is the following. After 50 psec of microcanonical molecular dynamics (MD) 
simulation performed at a temperature $T_{eq}$,  the protein is cooled down
by adding a linear dissipation term to the force acting on surface atoms,
that is, those belonging to amino-acids with more than 25~\AA$^2$ of
solvent accessible surface area. This represents nearly 40\% of the amino-acid residues,
for all proteins considered in the present study. 
The viscous friction coefficient $\g$ is set to 2 psec$^{-1}$, a typical value for protein atoms
in a water environment~\cite{Straub:00}. 
Hereafter, the equilibration energies considered are in the range 
$k_{\scs B} T_{eq}=2-20$ kcal/mol, 
that is, of the order of, e.g., the energy release of ATP hydrolysis.
With such initial conditions, energy in the system remains high 
for a period of time long enough so that localization can occur.

\begin{figure}[t!]
\vskip -4.0cm
\includegraphics[width=12.0 truecm]{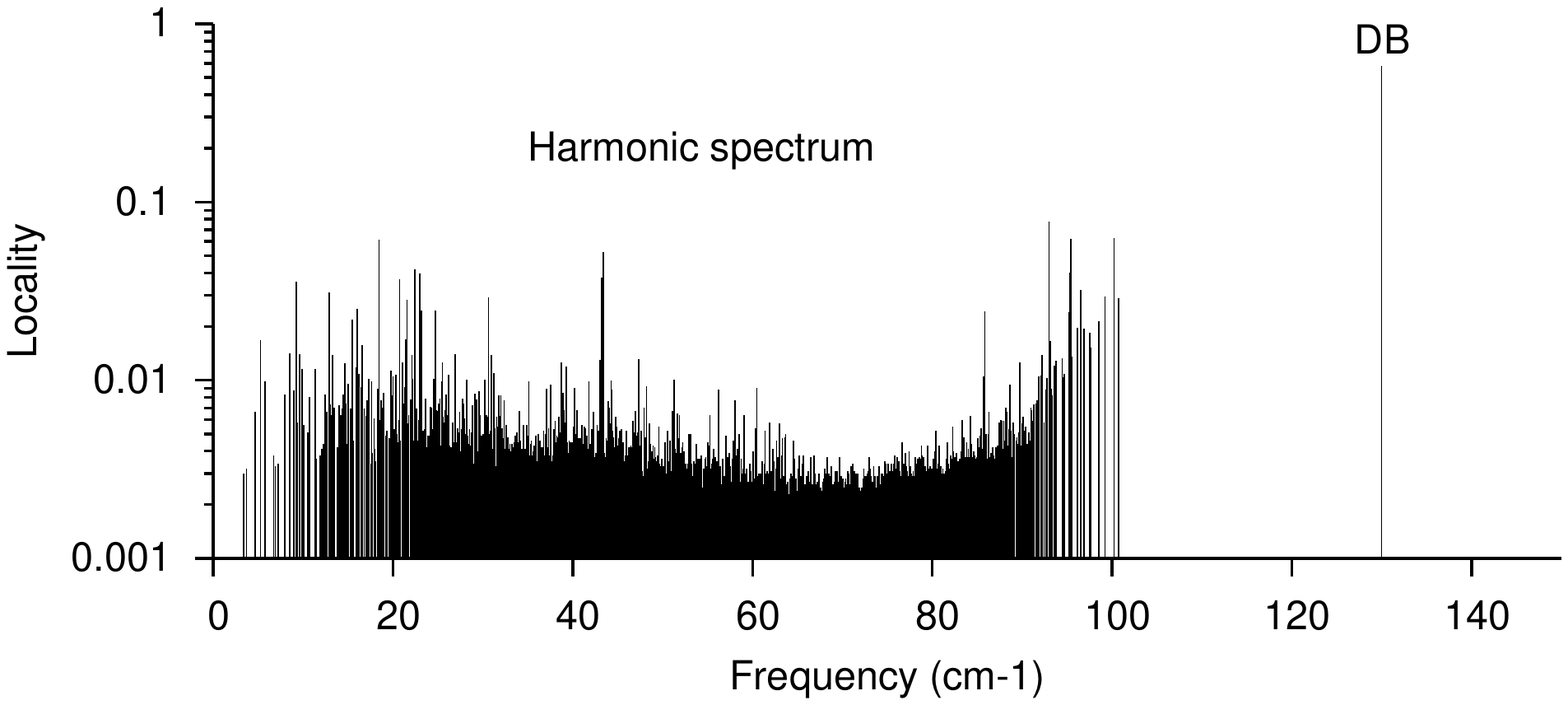}
\vskip -5mm
\caption{\small \label{spectrum} 
Locality of citrate synthase harmonic modes, as a function of their frequencies,
together with the locality and frequency of a discrete breather (DB).
}
\vskip -5mm
\end{figure}

In Fig.~\ref{etot}, we show the energy of dimeric citrate synthase (PDB code 1IXE)
as a function of time, as well as the energy of two amino-acids
of monomer A, Thr 208 and Ala 209. After $t=20$ psec 
and a few large fluctuations, a DB centered at Thr 208 forms.
At $t=200$ psec, more than 80\% of the total energy is located there.
Note the slow decay of the total energy after $t=$100 psec
and the periodic energy exchanges of Thr 208 with Ala 209, another among the few amino-acids 
involved in the DB.
Note also that at $t=20$ psec the energy of Thr 208 is higher than at $t=$0,
that is, when the friction was turned on, a clear-cut demonstration of the
known tendency of DBs to harvest energy from lower-energy excitations~\cite{Flach:1998fj}.
In order to check that the phenomenon shown in Fig.~\ref{etot} is
indeed the spontaneous localization of a DB,
we switched off the friction at $t=200$ psec and performed 100 more psec of
microcanonical MD simulation. Then,  we projected  the latter trajectory on 
the first eigenvector of the corresponding velocity-covariance  matrix,
which gives the pattern of correlated atomic velocities
involved in the DB.
The Fourier transform of such a projection yields an accurate value 
for the DB frequency, while the spectral line-width provides information
on the DB stability over the 100 psec analysis time-span.

\begin{figure}[t!]
\vskip -7.0cm
\hskip -1cm
\includegraphics[width=9.5 truecm]{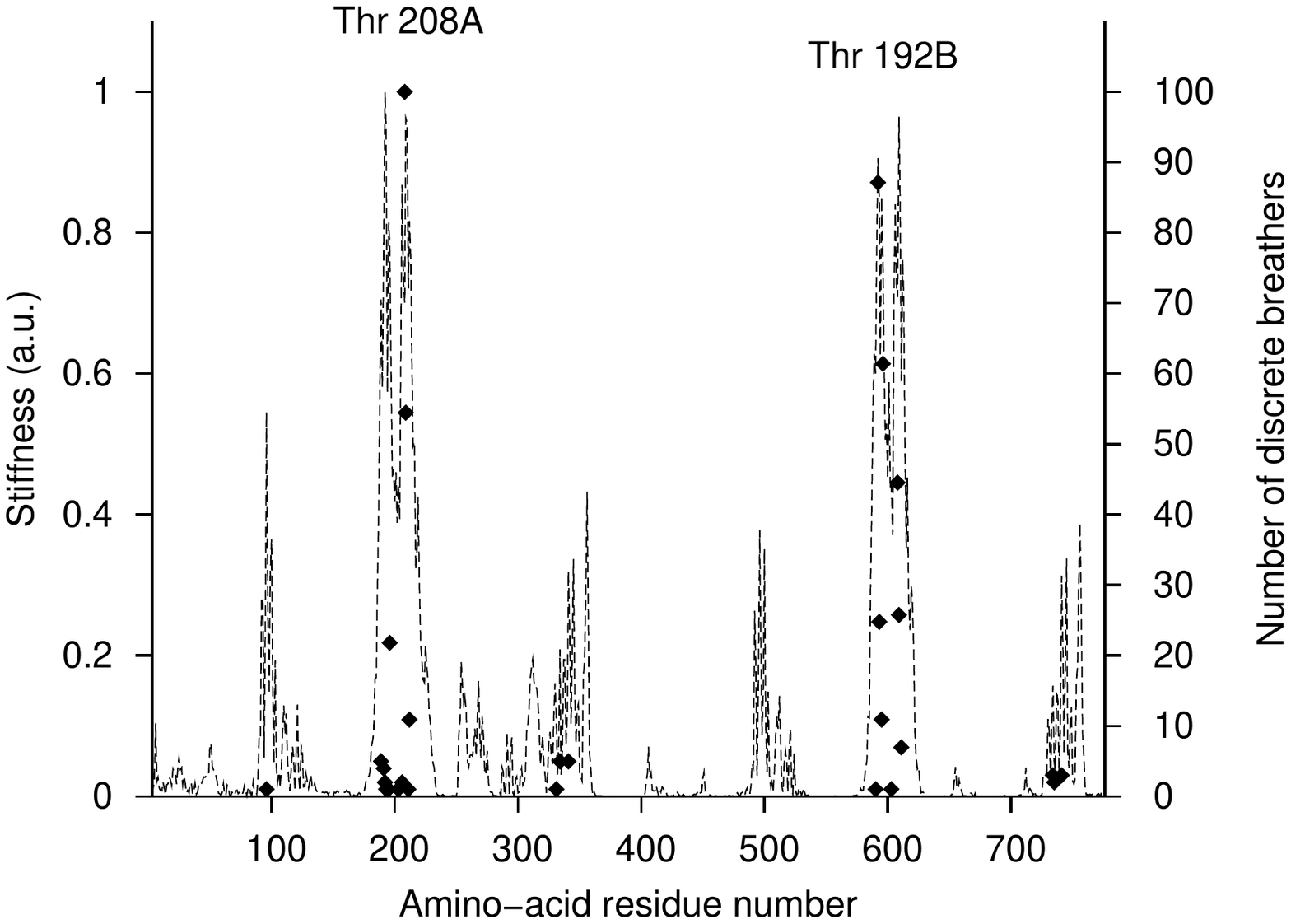}
\vskip -10mm
\caption{\small \label{DB}
Stiffness of dimeric citrate synthase as a function of residue number
(dashed line).  The number of DBs found at a given site
out of 500 instances  is also reported (black diamonds, right y-axis).}
\vskip -5mm
\end{figure}
In Fig.~\ref{spectrum} we report the harmonic spectrum of citrate synthase 
as well as the DB frequency
as functions of a locality measure. The latter is defined 
as $L_{k}= \sum_{i,\alpha} [\xi_{i\alpha}^k]^4/
[\sum_{i,\alpha} [\xi_{i\alpha}^k]^2]^2$, where $\xi_{i\alpha}^k$
is the $\alpha$ $ (x,y,z)$ coordinate of the $i$-th atom in the
$k$-th displacement pattern (normalized eigenvector, DB).
As expected, the DB frequency (130 cm$^{-1}$) lies above the 
highest frequency of the harmonic spectrum (101 cm$^{-1}$). Moreover,
the corresponding  spatial pattern is much more localized than any of the harmonic modes 
(note the logarithmic scale).

Starting from random initial conditions, we obtained 500 stable DBs following
the above-outlined protocol.
Although in many cases several DBs emerged, we decided to retain 
only the runs where a single DB catched most of the system energy,
and more energy than the average amount per site at $t=0$.
In Fig.~\ref{DB} we report the  number of DBs found at each site. 
The largest fraction (20 \%) of these highly energetic DBs formed
at Thr 208 in monomer A, but we also observed DBs  at 27 other sites, 
noteworthy at Thr 192 of monomer B (18\%).
Note also that, although the studied protein
is a dimer, that is, with a an approximate but clear two-fold structural symmetry, 
the probability to observe
a DB at a given site varies from one monomer to the other, indicating
that the localization dynamics is rather sensitive to small changes in the local environment.
As shown in Fig.~\ref{DB}, this probability is higher in the stiffest parts
of the protein scaffold, as measured through an indicator of local stiffness $s_{i}$.
For amino-acid $i$, the latter is defined as:
\begin{equation}
\label{Stiffness}
  s_i=\frac{1}{\mathcal{N}_i} 
       \sum_{j,\alpha} \sum_{k \in \mathcal{S}} [\xi_{j\alpha}^k]^2 \theta(R_{c}-d^0_{ij})
\end{equation}
where $\mathcal{N}_i = \sum_{j} \theta(R_{c}-d^0_{ij})$ is the number of neighbors of 
the $i$-th residue and $\theta(x)$ is the Heaviside step function. The second sum is
over the set $\mathcal{S}$ of the ten highest frequency harmonic modes. 
The averaging over the $\mathcal{N}_i$ neighbors slightly smoothes mode contributions and helps
underlining the fact that in each monomer of citrate synthase there
is a stretch of nearly fourty consecutive amino-acids (residues 185-225) with a
remarkably stiff environment, deeply buried at
the interface between the two monomers.
This is obviously where most DBs tend to emerge.
Note, however, that the relationship between high-frequency harmonic modes
and spontaneous energy localization is not a straightforward one: 
for instance, DBs were observed only a couple of times 
at the site the most involved in the highest frequency normal mode, namely,
Ser 213. As a matter of fact, as suggested by the large energy fluctuations
observed at site Thr 208 before the DB shown in Fig.~\ref{etot} springs up,
a competition between {\em potential} DBs is likely to occur, with possible 
weak-to-strong energy transfers, before a given site is occupied by a stable mode.

\begin{figure}[b]
\vskip -3.5cm
\hskip -5mm
\includegraphics[width=9.0 truecm]{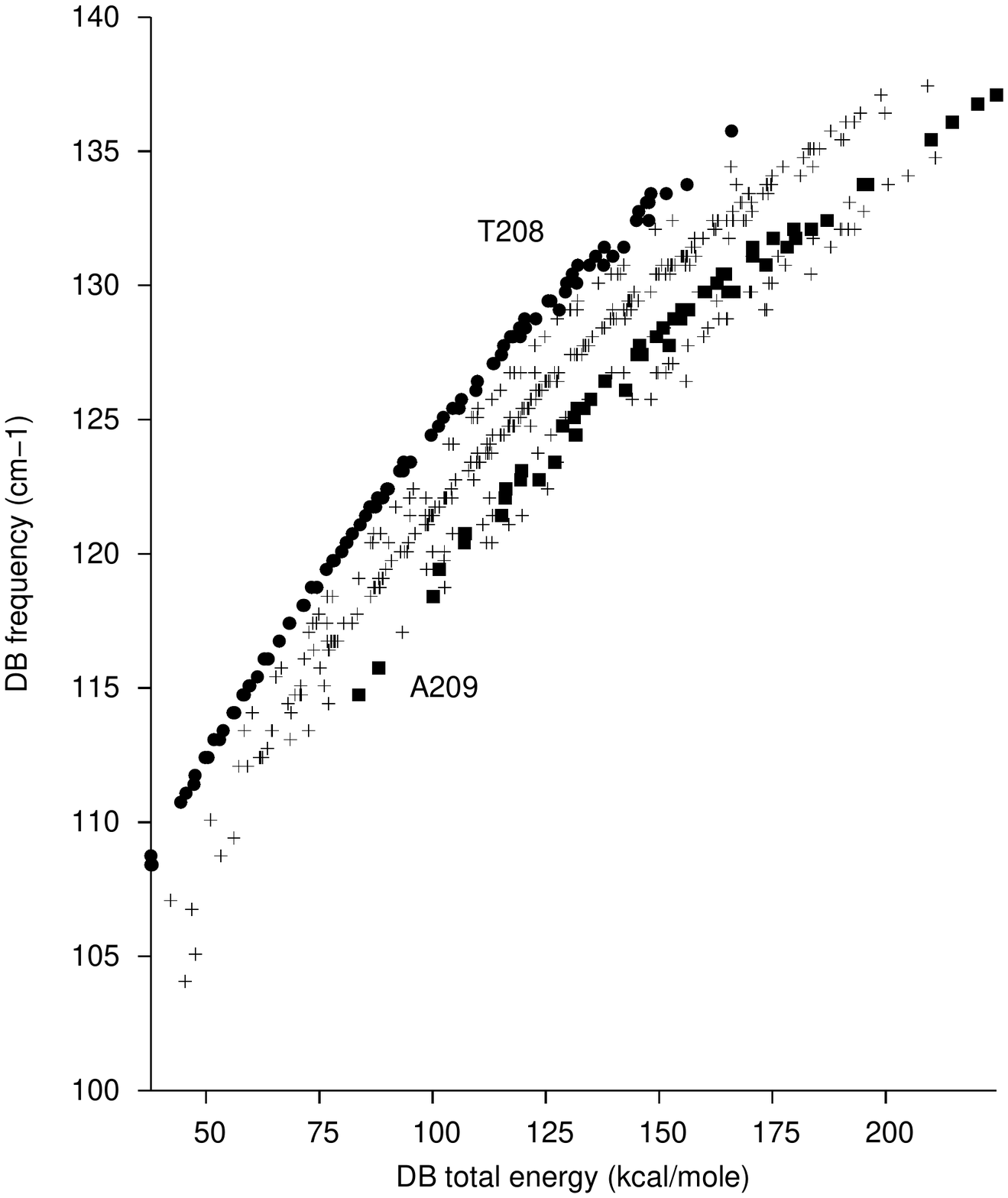}
\vskip -10mm
\caption{\small \label{enefreq}
DB frequencies in citrate synthase as a function of their energy (pluses).
The cases of Threonine 208 (filled circles) and Alanine 209 (filled squares)
are highlighted. Using our protocol, no DB with an energy
lower than 37.8 kcal/mole was observed, out of a total of 500 cases.
}
\vskip -5mm
\end{figure}

In lattice systems sites are obviously equivalent. Here, as shown in Fig.~\ref{enefreq},
the energy-frequency relationship  is site-dependent. 
Furthermore, the probability for a DB to localize at a given site depends in a non-obvious 
fashion  upon the energy it needs to reach a given frequency at that location.
While most DBs emerge at Thr 208, i.e. the site where the least energy is required
for a given frequency, many DBs are also observed at Ala 209 in monomer A, one 
of the sites that demands more energy. In more than one dimension one expects DBs to appear 
only above a characteristic energy~\cite{Kastner:2004yq,Flach:1998fj}. Hence,
our results hint at a strong site-dependence of such
energy threshold, non-trivially related to local structural properties.
To shed light on this intriguing feature, a detailed characterization of the small-ampitude side of 
the DB energy-amplitude curves at different sites is currently under way.

In the following step, we looked for DBs in other proteins, both dimeric and monomeric.
For small proteins like HIV-1 protease (PDB code 1A30), a dimeric $2 \times 99$ 
amino-acids enzyme, no DB could be obtained.
This is likely to be due to the fact that in small proteins too many
amino-acids are in direct interaction with a site where energy dissipation occurs. 
This means that small proteins may require more detailed models, 
like all-atom schemes, where cutoff values of the order of 5~\AA~ 
are customary~\cite{Elnemo1,Nicolay:06,Phillips:07}.

\begin{figure}[t!]
\vskip -7.0cm
\hskip -10mm
\includegraphics[width=9.5 truecm]{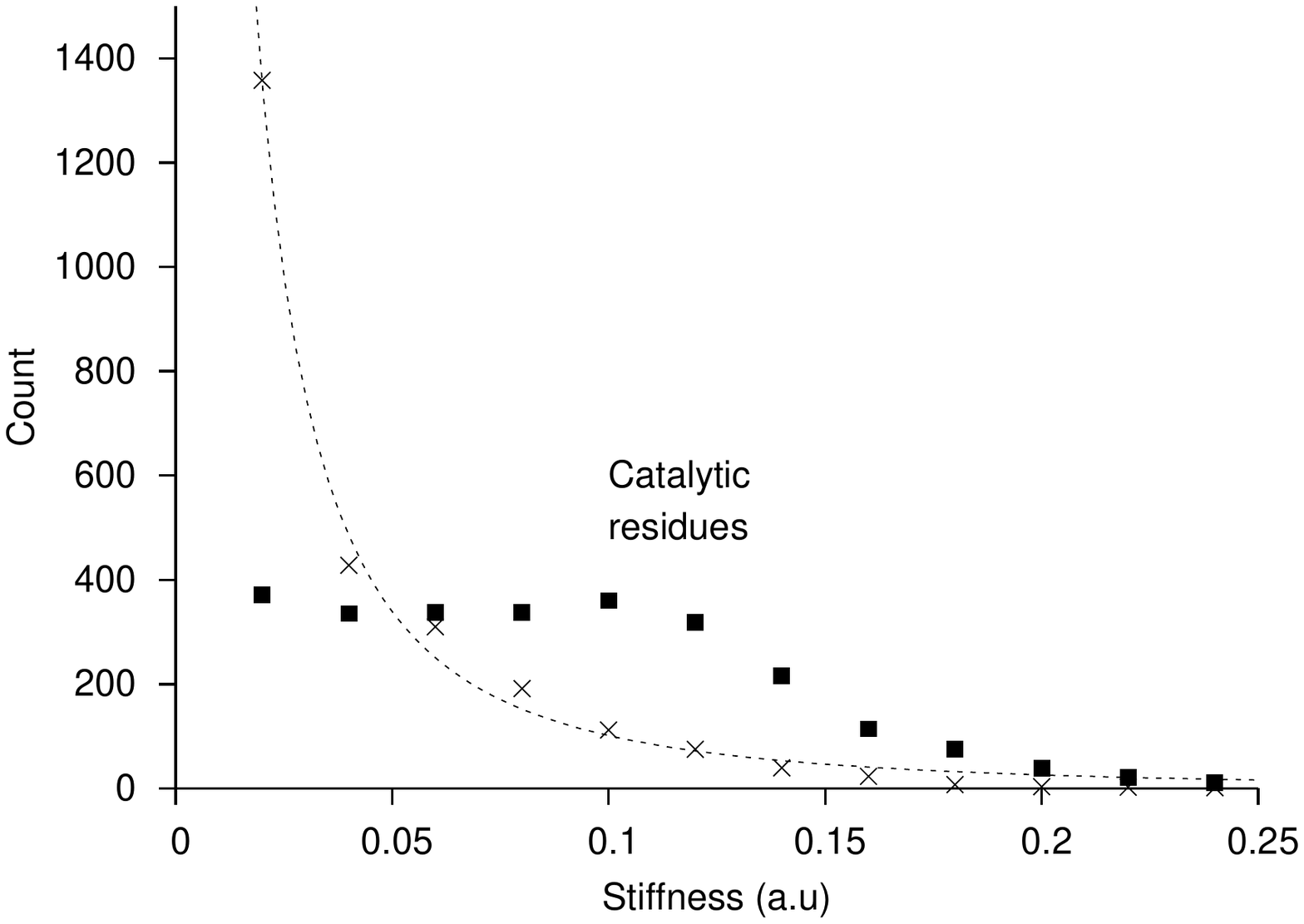}
\vskip -5mm
\caption{\small \label{Stats}
Stiffness of the environment of amino-acid residues involved in 
enzymatic activity (black squares),
compared to that of amino-acids of same chemical type (crosses) randomly chosen
within the same set of enzyme structures.
The broken line is only a guide for the eye.
}
\vskip -5mm
\end{figure}
In the case of aconitase (PDB code 1FGH), a monomeric 753 amino-acids enzyme, 
and alkaline phosphatase (PDB code 1ALK), a dimeric $2 \times 499$ amino-acids enzyme,
DBs prove nearly as easy to generate than in the case of 
citrate synthase. However, for proteins of similar sizes, the probability
of similar events turns out to vary significantly from a protein to another.  
For instance, in the cases of  phospholipase D (PDB code 1V0Y), a monomeric 504 amino-acids enzyme,
and isoamylase (PDB code 1BF2), a monomeric 750 amino-acids enzyme,
out of 100 cooling MD simulations, only 8 and 5 DBs were obtained, respectively,
in contrast to citrate-synthase, where our success rate is over 50 \%.
This points to the intriguing conclusion that not only DBs in proteins are site-selective,
but also appear to be non-trivially fold-selective.

In all the analyzed structures, spontaneous localization of energy  occurs
in the stiffest parts of the structure. Thus, we turn now to examine the relationship
between protein stiffness and function. Following the hypotheses
that enzymatic activity may require some kind of energy storage and that
DBs may play a role in the process, we computed high-frequency normal modes
for a set of 833 enzymes from the 2.1.11 version of the Catalytic Site Atlas~\cite{CSA}. 
Then, we determined the stiffness of each amino-acid known to be involved in enzymatic
activity according to~\eqref{Stiffness}. As a comparison, we also determined
stiffnesses of amino-acids of the same chemical type, but picked at random
among those not known to be involved in enzymatic activity.
As shown in Fig.~\ref{Stats}, catalytic amino-acids tend to be located
in stiffer parts of enzyme structures, in agreement with our hypotheses.
This is not an obvious result, since for the sake of catalytic activity
amino-acids have to interact with enzyme substrates, that is, to be accessible
to them. Such a trend has already been noticed in other studies.
Noteworthy, using
the ease of displacing any given amino-acid residue with respect to the others
as a stiffness measure, it was shown that 
roughly 80 \% of the catalytic residues are located in stiff parts
of enzyme structures~\cite{Lavery:07}. 
In a more indirect way, it was also remarked that global hinge centers
colocalize with catalytic sites in more than 70 \% of enzymes~\cite{Bahar:05}.
So, stiff parts may play a role of pivot, allowing for
accurate large-amplitude conformational changes of enzymes upon substrate binding.

What our results further suggest is that stiff parts
of enzyme structures may also play another major role in enzyme function, namely by
allowing for an active role of nonlinear localized modes in energy storage 
during the catalytic process.

Y-.H.S. wishes to thank M. Peyrard and T. Dauxois for an invitation to talk
at a training school held in Les Houches~\cite{DB:04},
where he was introduced to the fascinating world of discrete breathers.

\bibliography{protdyn}

\end{document}